\documentclass{elsart}
\usepackage{latexsym,amsmath,tabularx,amssymb,bm}
\usepackage{epsfig}
\usepackage{citesort}

\long\def\comment#1{ }

\usepackage{graphicx}
\usepackage{amssymb}
\usepackage{makeidx}
\usepackage{graphicx}
\usepackage{multicol}

\def\simge{\mathrel{%
    \rlap{\raise 0.511ex \hbox{$>$}}{\lower 0.511ex \hbox{$\sim$}}}}
\def\simle{\mathrel{
    \rlap{\raise 0.511ex \hbox{$<$}}{\lower 0.511ex \hbox{$\sim$}}}}

\newcommand \beq{\begin{eqnarray}}
\newcommand \eeq{\end{eqnarray}}

\begin{document}
\begin{frontmatter}

\parbox[]{16.6cm}{ \begin{center}

\title{Some non-renormalization theorems in Curci-Ferrari model}

\author[montevideo]{Nicol\'as Wschebor\thanksref{email}}

\address[montevideo]{ Instituto de F\'{\i}sica, Facultad de Ingenier\'{\i}a, Universidad de la Rep\'ublica, 
J.H.y Reissig 565, 11000
  Montevideo, Uruguay}

\thanks[email]{email: nicws@fing.edu.uy}

\date{\today}
\vspace{0.8cm}
\begin{abstract}
In the present letter, a particular form of Slavnov-Taylor identities for the Curci-Ferrari model is deduced.
This model consist of Yang-Mills theory in a particular non-linear covariant gauge, supplemented with mass terms
for gluons and ghosts. It can be used as a regularization for the Yang-Mills theory preserving simple
Slavnov-Taylor identities. Employing these identities two non-renormalization theorems are proved that reduce
the number of independent renormalization factors from five to three. These new relations are verified by
comparing to the already known three-loops renormalization factors.
These relations include, as a particular case, the corresponding known identities in Yang-Mills theory
in Landau gauge.
\end{abstract}
\end{center}}

\end{frontmatter}
\newpage

\section{Introduction}
In last years important progress has been made in the understanding of the non-perturbative sector of Yang-Mills
theory and QCD. The main reason for that comes from extensive Monte-Carlo simulations that have made possible
the calculation of different non-perturbative quantities. Lattice simulations have so opened the possibility to
test non-perturbative analytical
approximations. Probably the most important results of this kind are the Yang-Mills correlators, mainly using Schwinger-Dyson
\cite{Alkofer00,Zwanziger01,Aguilar01,Zwanziger02,Alkofer02b,Fischer02,Fischer02b,Aguilar02,Zwanziger03,Alkofer03,Bloch03,Schleifenbaum04,Alkofer04,Aguilar04}
and Non-Perturbative Renormalization Group \cite{Pawlowski03,Fischer04} equations and which have been compared
quite successfully with lattice results
\cite{Becirevic99b,Becirevic99,Bakeev03,Cucchieri03,Cucchieri04,Sternbeck05,Boucaud05}. These gauge-dependent objects have been
calculated in a variety of gauges (see, for example, \cite{Bornyakov03,Alkofer03b,Szczepaniak03,Zwanziger03b}) but most results
have been obtained in Landau gauge.

These results indicate that analytical calculations with relatively simple
approximations have a chance to be controlled not only in the ultraviolet sector of the theory but also in the
infrared, at least in Landau gauge. Contrarily to what happens in perturbation theory, the strong coupling does not
seem to diverge in the infrared at a finite momentum scale. Moreover, even if results depend on the scheme and on the details
of the approximations used, different analyses seem to agree on the fact that many naturally defined Yang-Mills couplings remain
bounded in the infrared. In the context of Landau gauge, the ghost-gluon coupling becomes particularly simple to analyze due to
the well-known non-renormalization theorem for this specific coupling \cite{Taylor71}. Many lattice results even seem to indicate
that this coupling goes to zero in the infrared \cite{Sternbeck05,Boucaud05}. On the same lines, gluon propagator do not have a
massless behaviour as obtained in perturbation theory but a massive one (Schwinger-Dyson indicates generally an infinite mass
\cite{Alkofer00,Zwanziger01,Zwanziger02,Fischer02b,Zwanziger03,Alkofer03} and lattices a finite one
\cite{Becirevic99b,Becirevic99,Sternbeck05,Boucaud05}).

This massive behaviour in the gluonic sector, with a frozen coupling in the infrared, reduces considerably infrared
instabilities of Yang-Mills theory. However, the ghost sector remains massless in Landau gauge at a non-perturbative
level and analytical or semi-analytical treatments must face typical difficulties of critical phenomena. In most
cases where quantitative results for critical systems have been achieved some sort of renormalization group
procedure has been necessary. The reason is that renormalization group equations only involve modes in a
relatively small shell of momenta. This is particularly true in the Wilsonian version of the renormalization group
\cite{Wilson73,Polchinski83,Wetterich92,Ellwanger93,Tetradis94,Morris94b,Morris94,Berges00} but in some sense is the
characteristic property of different kinds of renormalization group equations. This must be contrasted with
Schwinger-Dyson equations where all momenta are coupled in a significant way. One manifestation of this is that
these equations require an explicit renormalization. To implement this renormalization process in a non-perturbative way
is far from trivial. This is a serious motivation in order to look at the construction of renormalization
group equations in Yang-Mills theory describing in a complete way correlation functions.

The ideal tool in this sense would be to have a Wilsonian flow equation for Yang-Mills theory. However, the standard
procedure \cite{Becchi03,Ellwanger94,Bonini93,Bonini93b,Bonini94b,Bonini94,Ellwanger95,Ellwanger96} does not
preserve simple Slavnov-Taylor identities \cite{Taylor71,Slavnov72} along the flow and an alternative formulation
that preserve gauge symmetry has turned to be extremely involved \cite{Morris99,Morris00,Arnone01,Arnone02,Morris05,Rosten05,Morris06}.
Another renormalization group given by Callan-Symanzik equation \cite{Callan70,Symanzik70} describes the flow of a
field theory in term of a renormalized mass parameter that regulates
completelly the theory in the infrared. It has poorer non-perturbative properties than Wilsonian flows, but is
much simpler.
In the Yang-Mills case, this requires a deformation of the theory in such a way that gluons and ghost acquire a mass, preserving
renormalizability and reasonably simple Slavnov-Taylor identities. The simplest way to do that is offered by the Curci-Ferrari 
model \cite{Curci76,Curci76b}. It corresponds to the Yang-Mills theory in a certain family of gauge fixings supplemented with a
particular mass terms for gluons and ghosts. If the gauge and mass parameters go to zero, Curci-Ferrari model reduces to the important
case of Yang-Mills theory in Landau gauge.
It may be useful to remember here that only the theory with zero mass parameter is
physically acceptable. In fact, it has been proved that only in this case the model verifies perturbative unitarity and gauge
parameter independence (see for example \cite{Curci76b,Ojima81,deBoer95}). This, however, does not excludes
the use of this model as an intermediate step regularization if at the end of the calculation the mass parameter is taken to be zero.

There is another motivation to consider the Curci-Ferrari model. The non-perturbative definition of a BRST symmetry
\cite{Becchi74,Becchi75,Tyutin75} has been elusive. In particular, Neuberger has constructed a lattice version of the BRST symmetry
\cite{Neuberger86} but, he also showed \cite{Neuberger86b} that averages of gauge invariant operators take the form in any finite
lattice of a zero over zero expression. This difficulty is closely related to the Gribov problem \cite{Gribov77} of non-perturbative
gauge fixing in non abelian gauge theories.
However, recently it has been proved \cite{Kalloniatis05} that the Curci-Ferrari model regularize this zero, gauge invariant
correlators acquiring a definite expression for any non-zero mass.
This allows for a non-perturbative construction of the BRST-invariant model on a lattice \cite{Ghiotti06}. This is radically different
to usual gauge theories where 
Neuberger's zero forbids such a construction \cite{Neuberger86b}.
In this way, Yang-Mills theory can be seen as the zero limit of the Curci-Ferrari model whose BRST symmetry has a non-perturbative
meaning. Moreover, the corresponding limit can be seen as the result of the renormalization group flow.

Finally, the Landau-gauge has more symmetries than Curci-Ferrari gauge and so is fixed under renormalization.
However, if one takes a zero gauge parameter along the flow, ghost remain massless and Neuberger's zero is not
regularized. Fortunately, the Landau gauge (where most of the gauge-fixed Monte-Carlo simulations have been performed)
seems to be an attractive renormalization group fixed point (see, for example, \cite{Ellwanger95}). If so, the zero
mass limit automatically pushes the theory to the Landau gauge in the infrared.

For all these reasons, the analysis of Curci-Ferrari model is interesting. However, even if multiplicatively
renormalizable, Curci-Ferrari model seemed to require five renormalization factors \cite{deBoer95}. In fact, a
claim of a reduced number of renormalization factors \cite{Periwal95}, turned out to be incorrect \cite{deBoer95}.
These renormalization factors can be taken to be those corresponding to the
renormalization of gluon and ghost fields, gauge parameter, coupling constant and mass and have been calculated up
to three loops. This is to be compared to usual linear covariant gauges where (without masses) three
renormalization factors are necessary or to the Landau gauge where only two are required. This complicates
considerably any analytical or semi-analytical analysis. In this letter, a perturbative analysis of the
renormalization procedure of the Curci-Ferrari model is performed showing that, in fact, only three renormalization
factors are required. Unexpectedly, some symmetries present in the model had not been fully taken into account up to now
and, in fact, the mass and coupling renormalization can be expressed in terms of the other three renormalization factors. These relations are generalizations of the already known Landau gauge non-renormalizations theorems for these quantities \cite{Taylor71,Dudal02}.
This result is only a consequence of the high level of symmetry of the model that may be very useful in non-perturbative approximations.

The outline of the letter is the following: First, the Curci-Ferrari model and its symmetries are reviewed.
Second, the corresponding Slavnov-Taylor identities and the associated consequences for the
renormalization factors are deduced. Finally some conclusions and perspectives are presented.

\section{The Curci-Ferrari model}

The Curci Ferrari model is a modification of Yang-Mills theory that includes masses for gluons and ghost preserving
a sort of BRST symmetry. It has also many other symmetries, one of them being an anti-BRST one.
Here the model without matter fields in a 4-dimensional euclidean space-time is considered, but it is straightforward to see
that the results to be proved are true also including matter fields or considering a Minkowski space-time.

The variations under BRST and anti-BRST of different fields are:
\begin{equation}
\label{symm}
\begin{array}{ll}
s A_\mu^a=D_\mu c^a,
&\hspace{1cm}\bar s A_\mu^a= D_\mu \bar c^a,\\
s c^a= -\frac{g_0}{2} f^{abc} c^bc^c,
&\hspace{1cm}\bar s c^a=-\frac{1}{\xi_0}\partial_\mu A_\mu^a-\frac{g_0}{2} f^{abc} \bar c^bc^c,\\
s \bar c^a =\frac{1}{\xi_0}\partial_\mu A_\mu^a -\frac{g_0}{2} f^{abc}\bar c^b c^c,
&\hspace{1cm}\bar s \bar c^a =-\frac{g_0}{2} f^{abc} \bar c^b \bar c^c.
\end{array}
\end{equation}
Here, $g_0$ is the bare gauge coupling and $\xi_0$ is the bare gauge parameter of the model.
$A_\mu$ is the gauge field, $c$ and $\bar c$ are ghost and antighosts respectively. $D_\mu c^a$ denotes
the ghost covariant derivative $\partial_\mu c^a +g_0 f^{abc}A_\mu^b c^c$ and similarly for antighost.

The most general renormalizable Lagrangian with these symmetries and the space time euclidean symmetry, global color
invariance and the conservation of the number of ghost take the form:
\begin{equation}
\mathcal{L}=\mathcal{L}_{YM}+\mathcal{L}_{GF}+\mathcal{L}_m.
\end{equation}
Here $\mathcal{L}_{YM}$ is the Yang-Mills lagrangian:
\begin{equation}
\mathcal{L}_{YM}=\frac{1}{4}F_{\mu\nu}^aF_{\mu\nu}^a,
\end{equation}
where $F_{\mu\nu}^a=\partial_\mu A_\nu^a-\partial_\nu A_\mu^a+g_0 f^{abc}A_\mu^b A_\nu^c$.
$\mathcal{L}_{GF}$ is the gauge fixing term, that includes a ghost sector. It takes the form:
\begin{equation}
\mathcal{L}_{GF}=\frac{1}{2}\partial_\mu \bar c^a D_\mu c^a
+\frac{1}{2}D_\mu \bar c^a \partial_\mu c^a+\frac{1}{2\xi_0}(\partial_\mu A_\mu^a)^2
-\xi_0\frac{g_0^2}{8}f^{abc}f^{ade}\bar c^bc^c\bar c^dc^e.
\end{equation}
Finally the mass term $\mathcal{L}_m$ takes the form:
\begin{equation}
\mathcal{L}_m=m_0^2\left(\frac{1}{2} A_\mu^aA_\mu^a+\xi_0 \bar c^a c^a\right),
\end{equation}
where $m_0$ is the bare mass.

It is important to observe that the action has also other linearly realized symmetries \cite{Delduc89}.
First, there is the discrete symmetry where $c^a\to \bar c^a$ and $\bar c^a\to -c^a$.
There are also two continuous symmetries. The first one has the infinitesimal form $c^a\to c^a +\epsilon \bar c^a$.
The second, independent, symmetry correspond to $\bar c^a\to \bar c^a +\epsilon c^a$. These symmetries, as BRST
and anti-BRST do not present anomalies if the corresponding massless model do not. This may be seen from the
fact that there is an explicit lattice regularization which respect all these symmetries \cite{Ghiotti06}.

BRST and anti-BRST transformations are not nilpotent in the Curci-Ferrari model even after imposing equations of
motion. However, one can deduce closed Slavnov-Taylor identities by exploiting equations of motion for ghost and
anti-ghost fields. Consider the $s^2$ and $\bar s^2$ operations:
\begin{align}
s^2 A_\mu^a&=\bar s^2 A_\mu^a=s^2 c^a=\bar s^2 \bar c^a=0, \nonumber\\
\bar s^2 c^a&=-\frac{1}{\xi_0}\partial_\mu(D_\mu\bar c^a)
+\frac{g_0}{2} f^{abc}\bar c^b \Big(-\frac{1}{\xi_0}\partial_\mu A_\mu^c
-\frac{g_0}{2} f^{cde}\bar c^d c^e \Big) +\frac{g_0^2}{4} f^{abc}f^{bde}\bar c^d \bar c^e c^c,
\nonumber\\
s^2 \bar c^a&=\frac{1}{\xi_0}\partial_\mu(D_\mu c^a)
-\frac{g_0}{2} f^{abc}\Big(\frac{1}{\xi_0}\partial_\mu A_\mu^b
-\frac{g_0}{2} f^{bde}\bar c^d \bar c^e \Big)c^c -\frac{g_0^2}{4} f^{abc} f^{cde} \bar c^b c^d c^e.
\end{align}
Before deducing Slavnov-Taylor identities for these symmetries it is useful to consider also 'mixed' variations for different fields:
\begin{align}
\bar s s A_\mu^a&= D_\mu \Big(-\frac{1}{\xi_0}\partial_\nu A_\nu^a
-\frac{g_0}{2} f^{abc}\bar c^b c^c\Big)
+g_0 f^{abc} (D_\mu \bar c^b) c^c, \nonumber\\
\bar s s A_\mu^a&=-s \bar s A_\mu^a, \nonumber \\
\bar s s c^a &= g_0 f^{abc} c^b \Big(-\frac{1}{\xi_0}\partial_\nu A_\nu^c
-\frac{g_0}{2} f^{cde}\bar c^d c^e\Big), \nonumber\\
s \bar s c^a&= -\frac{1}{\xi_0}\partial_\mu D_\mu c^a
-\frac{g_0^2}{4} f^{abc}f^{cde} \bar c^b c^d c^e
-\frac{g_0}{2} f^{abc} \Big(\frac{1}{\xi_0}\partial_\nu A_\nu^b
-\frac{g_0}{2} f^{bde}\bar c^d c^e\Big) c^c, \nonumber\\
\bar s  s \bar c^a&=\frac{1}{\xi_0}\partial_\mu D_\mu \bar c^a
+\frac{g_0^2}{4} f^{abc}f^{bde} \bar c^d \bar c^e c^c
+\frac{g_0}{2} f^{abc} \bar c^b \Big(- \frac{1}{\xi_0}\partial_\nu A_\nu^c
-\frac{g_0}{2} f^{cde}\bar c^d c^e\Big), \nonumber\\
s \bar s \bar c^a&=g_0 f^{abc} \bar c^b \Big(\frac{1}{\xi_0}\partial_\nu A_\nu^c
-\frac{g_0}{2} f^{cde}\bar c^d c^e\Big).
\end{align}
In order to deduce Slavnov-Taylor identities for these symmetries, it is necessary to introduce sources for the variations
of different fields under BRST and anti-BRST symmetries.
Because the operators are not nilpotent, one should include also sources for the $s^2$ and $\bar s^2$ variations.
Moreover, the introduction of sources for the variations of fundamental fields under $s \bar s$, should be necessary a priori
(see, for exemple, \cite{Alvarez-Gaume81}) because variations under anti-BRST transformations are not BRST invariant as just seen.
However, as will be seen below, almost all these double variations can be expressed in terms of fundamental fields and simple variations
by exploiting ghost and antighost equations of motion. The only variation that can not be obtained in this way is $s \bar s A_\mu^a$.
Even this particular variation will not be needed in the present letter where Slavnov-Taylor identities are only considered at $\bar K=0$.
Let's consider the generating functional:
\begin{align}
\label{pathintegral}
\exp(W[J,\chi,\bar\chi,K,L,\bar K,\bar L])
&=\int D(A,c,\bar c) \exp\Bigg\{\int d^4x\Big(-\mathcal{L}+
J_\mu^a A_\mu^a+\chi^a\bar c^a+\bar\chi^a c^a \nonumber \\
&+K_\mu^a sA_\mu^a+L^a sc^a+\bar K^a_\mu \bar s A^a_\mu+\bar L^a \bar s \bar c^a
+M^a(-g_0/2 f^{abc}\bar c^b c^c)
\nonumber \\
&-\frac{1}{2\xi_0}M^aM^a+\frac{2}{\xi_0}\bar L^a L^a+\Bar K_\mu^a K_\mu^a\Big)\Bigg\}.
\end{align}
Notice that sources for composite operators are coupled non-linearly. As will be seen bellow this is necessary
in order to preserve multiplicative renormalizability. Table 1 contains dimensionalities
and ghost numbers for different fields and sources.
\begin{table}
\begin{center}
\begin{tabular}{|c|c|c|c|c|c|c|c|c|}
\hline 
\ \ Field/Source\ \  & $A$ & $c$ &$\bar c$& $K$ &$\bar K$& $L$ &$\bar L$& $M$ \\
\hline 
Dimension            & 1   &  1  &    1   &  2  &    2   &  2  &   2    & 2\\ \hline
Ghost number         & 0   &  1  &   -1   &  -1 &    1   & -2  &   2    & 0 \\ \hline
\end{tabular}
\caption{\it Canonical dimensionalities and ghost number of different fields and sources.}
\end{center}
\label{Table}
\end{table}
As usual, the Slavnov-Taylor identity associated to BRST symmetry is obtained by performing the change of variables in the
functional integral $\varphi\to \varphi
+\delta s\varphi$ ($\varphi$ being all the fundamental fields and $\delta$ a constant
grassmanian parameter). One obtains for $\bar K=0$:
\begin{equation}
\label{toST}
\int d^4x \Big\langle J_\mu^a s A_\mu^a
-\bar \chi^a s c^a - \chi^a s \bar c^a
+ \bar L^a s \bar s \bar c^a -\frac{g_0}{2}M^a f^{abc} s\big(\bar c^b c^c\big)
\Big\rangle\Big|_{\bar K=0}=0.
\end{equation}
In order to close the equation we need expressions for $\langle s \bar c^a\rangle$,
$\langle s \bar s \bar c^a\rangle$, and $f^{abc}\langle s (\bar c^b c^c)\rangle$. To obtain them one can use the equations of
motion for ghost and anti-ghost fields. The ghost equation of motion is obtained by performing in the generating functional the
infinitesimal change of variables
$c^a(x)\to c^a(x)+\epsilon^a(x)$ giving:
\begin{equation}
\bar \chi^a+\frac{g_0}{2}f^{abc}\bar c^b M^c + g_0 f^{abc} c^b L^c - D_\mu K_\mu^a
=\Big\langle
-\frac{1}{2}D_\mu \partial_\mu \bar c^a-\frac{1}{2}\partial_\mu D_\mu \bar c^a
+\frac{\xi_0}{4}g_0^2 f^{abc}f^{cde}\bar c^b \bar c^d c^e\Big\rangle
+m_0^2\xi_0 \bar c^a.
\end{equation}
Here the same notation has been used for the average of fundamental fields and for the fields themselves.
Along the same lines one obtains:
\begin{equation}
\chi^a+\frac{g_0}{2}f^{abc}M^b c^c + g_0 f^{abc} \bar c^b \bar L^c - D_\mu \bar K_\mu^a
=\Big\langle
\frac{1}{2}D_\mu \partial_\mu c^a+\frac{1}{2}\partial_\mu D_\mu c^a
-\frac{\xi_0}{4}g_0^2 f^{abc}f^{bde} c^c \bar c^d c^e\Big\rangle
-m_0^2\xi_0 c^a.
\end{equation}
This implies the following relations:
\begin{align}
\label{relations}
\xi_0 \langle \bar s^2 c^a\rangle&=\bar \chi^a+\frac{g_0}{2}f^{abc}\bar c^b M^c + g_0 f^{abc} c^b L^c - D_\mu K_\mu^a-m_0^2\xi_0 \bar c^a,
\nonumber\\
\xi_0 \langle s^2 \bar c^a \rangle&=\chi^a+\frac{g_0}{2}f^{abc}c^b M^c  + g_0 f^{abc} \bar c^b \bar L^c - D_\mu \bar K_\mu^a+m_0^2\xi c^a,
\nonumber \\
-\xi_0\langle (s\bar s+\bar s s)\bar c^a \rangle&=\bar \chi^a+\frac{g_0}{2}f^{abc}\bar c^b M^c + g_0 f^{abc} c^b L^c - D_\mu K_\mu^a
-m_0^2\xi_0 \bar c^a, \nonumber \\
\xi_0\langle s \bar s \bar c^a\rangle&=2\Big(-\partial_\mu \langle \bar s A_\mu^a \rangle-\bar \chi^a-\frac{g_0}{2}f^{abc}\bar c^b M^c
- g_0 f^{abc} c^b L^c+ D_\mu K_\mu^a+m_0^2\xi_0 \bar c^a \Big), \nonumber\\
-\xi_0 \langle (s \bar s +\bar s s) c^a\rangle &=\chi^a+\frac{g_0}{2}f^{abc}c^b M^c  + g_0 f^{abc} \bar c^b \bar L^c - D_\mu \bar
K_\mu^a+m_0^2\xi_0 c^a,
\nonumber\\
\xi_0\langle \bar s s c^a\rangle &=2\Big(\partial_\mu \langle s A_\mu^a\rangle-\chi^a-\frac{g_0}{2}f^{abc} c^b M^c
- g_0 f^{abc} \bar c^b \bar L^c + D_\mu \bar K_\mu^a-m_0^2\xi_0 c^a\Big).
\end{align}
Finally, the last remaining variation can be obtained noting that:
\begin{equation}
\label{relation2}
\Big\langle s\big(-\frac{g_0}{2}f^{abc}\bar c^b c^c\big) \Big\rangle=
\Big\langle s\big(\frac{1}{\xi_0}\partial_\mu A_\mu^a-\frac{g_0}{2}f^{abc}\bar c^b c^c
-\frac{1}{\xi_0}\partial_\mu A_\mu^a\big) \Big\rangle=
\Big\langle s^2 \bar c^a
-\frac{1}{\xi_0}\partial_\mu s A_\mu^a\big) \Big\rangle.
\end{equation}
Please observe an important non-trivial fact, all double variations (aside $s\bar s A_\mu^a$) have been expressed in terms
of operators whose sources are included in eq.~(\ref{pathintegral}).

Before considering renormalization properties of the model, it is important to realize that under the
infinitesimal linear symmetry
\begin{equation}
\label{linsym1}
c^a\to c^a+\epsilon \bar c^a,
\end{equation}
composite operators mix in the following way:
\begin{align}
&s A_\mu^a \to s A_\mu^a +\epsilon \bar s A_\mu^a \nonumber \\
&s c^a \to s c^a -\epsilon g_0 f^{abc}\bar c^b c^c \nonumber \\
&s \Big( -\frac{g_0}{2}f^{abc}\bar c^b c^c\Big)\to
s \Big( -\frac{g_0}{2}f^{abc}\bar c^b c^c\Big)
-\epsilon \frac{g_0}{2}f^{abc}\bar c^b \bar c^c=\epsilon \bar s\bar c^a
\end{align}
all the other being invariant. This is equivalent to saying that sources for composite operators vary linearly in the following way:
\begin{align}
&K_\mu^a \to K_\mu^a +\epsilon \bar K_\mu^a \nonumber  \\
&\bar L^a \to \bar L^a +\epsilon M^a \nonumber \\
&M^a \to M^a+2 \epsilon L^a
\end{align}
On the same lines, one deduce that under the infinitesimal symmetry
\begin{equation}
\label{linsym2}
\bar c^a\to \bar c^a+\epsilon c^a,
\end{equation}
sources for composite operators vary linearly also:
\begin{align}
&\bar K_\mu^a \to \bar K_\mu^a +\epsilon K_\mu^a \nonumber \\
&L^a \to L^a +\epsilon M^a \nonumber \\
&M^a \to M^a+2 \epsilon \bar L^a
\end{align}
Please observe that terms non-linear in sources are invariant under these transformations.

\section{Non-renormalization theorems}

In this section, the renormalization properties of the CF model implied by ST identities previously presented
is considered.
By substituting expressions (\ref{relations}) and (\ref{relation2}) in (\ref{toST}) and expressing the results in terms of the
effective action one arrives to: 
\begin{align}
\label{STBRST}
&\int d^4x\Bigg\{
\frac{\delta \Gamma}{\delta A_\mu^a}\frac{\delta \Gamma}{\delta K_\mu^a}
+\frac{\delta \Gamma}{\delta c^a}\frac{\delta \Gamma}{\delta L^a}
+\frac{\delta \Gamma}{\delta \bar c^a}\Bigg(-\frac{1}{\xi_0}
\partial_\mu A_\mu^a+\frac{\delta \Gamma}{\delta M^a}\Bigg) \nonumber\\
&-\frac{1}{\xi_0} M^a \partial_\mu\Bigg( \frac{\delta \Gamma}{\delta K_\mu^a} \Bigg) 
-\frac{2}{\xi_0} \bar L^a \Bigg( \partial_\mu\Bigg(\frac{\delta \Gamma}{\delta \bar K_\mu^a}\Bigg)
-g_0 f^{abc}c^b L^c + D_\mu K^a_\mu +m_0^2 \xi_0 \bar c^a \Bigg)
\Bigg\}\Bigg|_{\bar K=0}=0.
\end{align}

The perturbative renormalizatibilty of CF model has been proven by considering five renormalization factors \cite{deBoer95}.
Here this proof is revisited by exploiting this particular form of the ST identity. The non-trivial use of equations of motion presented
before will imply that only three renormalization factors are required.
To prove that, I follow the standard procedure \cite{Weinberg96,Nakanishi90} of considering terms that can diverge by power counting and
constraining them iteratively. In a loop expansion, suppose that all divergences have been renormalized at order $n$.
Diverging terms that appear in the order
$n+1$ in the effective action, have couplings with positive or zero dimension.
Let's call them $\Delta \Gamma_{div}$, and take an infinitesimal constant, $\epsilon$.  
If one calls $\Gamma_{div}=S+\epsilon \Delta \Gamma_{div}$, then, in dimensions lower than four, the most general
form for this functional at order $n+1$, takes the form:
\begin{align}
\label{previous1}
\Gamma_{div}[A,c,\bar c,K,\bar K,L,\bar L,M]&=\hat\Gamma[A,c,\bar c]
-\int d^dx\Bigg\{\frac{2}{\xi} \bar L^a L^a -\frac{1}{2\xi} M^a M^a+\gamma \bar K_\mu^a K_\mu^a \nonumber \\
&+K_\mu^a \tilde s A_\mu^a +\bar K_\mu^a \tilde{ \bar{ s}} A_\mu^a
+L^a \tilde s c^a +\bar L^a \tilde{\bar{s}}\bar c^a+M^a \Bigg(
-\frac{1}{\xi_0}\partial_\mu A_\mu^a +\tilde s \bar c^a\Bigg)
\Bigg\}.
\end{align}
where, for the moment, $\tilde s A_\mu^a$, $\tilde{ \bar{ s}} A_\mu^a$,
$\tilde s c^a$, $\tilde{\bar{s}}\bar c^a$ and $\tilde s \bar c^a$ denote arbitrary operators of dimension two
with the same transformations under linear symmetries as the corresponding bare expressions. Here the coefficients
of $\bar L^a L^a$ and $M^a M^a$ are related by the linear symmetry (\ref{linsym1}).

Substituting these expression into the Slavnov-Taylor identity (\ref{STBRST}), one can analyse first the particular
case corresponding to $\bar L=M=0$. This gives the three relations:
\begin{align}
&\tilde s^2 A_\mu^a=0, \label{nil1}\\
&\tilde s^2 c^a=0, \label{nil2}\\
&\tilde s \hat \Gamma =0. \label{sgamma}
\end{align}
(see, for exemple, \cite{Weinberg96} for details of an analogous deduction in a similar situation).
Now, the most general operators of dimension two, respecting Lorentz invariance, global color invariance, symmetries
(\ref{linsym1}) and (\ref{linsym2}) and nilpotecy (\ref{nil1},\ref{nil2}) are
\begin{align}
\label{previous2}
\tilde s A_\mu^a &= Z \partial_\mu c^a +g f^{abc} A_\mu^b c^c, \nonumber\\
\tilde{ \bar{ s}} A_\mu^a &=Z \partial_\mu \bar c^a + g f^{abc}
A_\mu^b \bar c^c, \nonumber\\
\tilde s c^a &= -\frac{g}{2}f^{abc} c^b c^c, \nonumber\\
\tilde{\bar{s}}\bar c^a &= -\frac{g}{2} f^{abc} c^b c^c, \nonumber\\
\tilde s \bar c^a&=\frac{1}{\xi_0} \partial_\mu A_\mu^a
-\frac{g}{2}f^{abc} \bar c^b c^c.
\end{align}
The equation linear in $M$ at $L=\bar L=K=0$ becomes then,
\begin{equation}
\tilde s^2 \bar c^a
+m_0^2 c^a=\frac{1}{\xi}\frac{\delta \hat \Gamma}{\delta \bar c^a}. \label{ecmovbarc}
\end{equation}
Then, imposing the two identities (\ref{ecmovbarc},\ref{sgamma}) one deduces that the most general expression for
$\hat \Gamma$ takes the form $\hat \Gamma=\Gamma_m+\Gamma_{GF}+\Gamma_{YM}$, where:
\begin{align}
\Gamma_m&=\int d^4x \frac{m_0^2}{Z_m} \Bigg\{ \frac{1}{2Z_A} A_\mu^aA_\mu^a+\frac{\xi_0 Z_\xi}{Z_A Z_c} \bar c^a c^a\Bigg\},
\nonumber \\
\Gamma_{GF}&=\int d^4x \Bigg\{ \frac{1}{2 Z_c} \partial_\mu \bar c^a\tilde D_\mu c^a+
\frac{1}{2 Z_c} \tilde D_\mu \bar c^a\partial_\mu c^a+\frac{1}{2\xi_0 Z_\xi}(\partial_\mu A_\mu^a)^2
-\frac{Z_\xi \xi_0 g_0^2}{8 Z_g^2 Z_A Z_c^2} f^{abc}f^{ade}\bar c^b c^c\bar c^d c^e\Bigg\}, \nonumber\\
\Gamma_{YM}&=\int d^4x \Bigg\{ \frac{1}{4 Z_A} \tilde F^a_{\mu\nu}\tilde F^a_{\mu\nu}\Bigg\}.
\end{align}
with
\begin{align}
\tilde D_\mu c^a&=\partial_\mu c^a + \frac{g_0}{Z_g \sqrt{Z_A}} f^{abc} A_\mu^b c^c, \nonumber\\
\tilde D_\mu \bar c^a&=\partial_\mu \bar c^a + \frac{g_0}{Z_g \sqrt{Z_A}} f^{abc} A_\mu^b \bar c^c, \nonumber\\
\tilde F^a_{\mu\nu}&=\partial_\mu A_\nu^a-\partial_\nu A_\mu^a+\frac{g_0}{Z_g \sqrt{Z_A}} f^{abc} A_\mu^bA_\nu^b.
\end{align}
These factors can be absorbed in the following renormalizations:
\begin{equation}
\label{renfact}
\begin{array}{ll}
A_{\mu}^a=\sqrt{Z_A} A_{R,\mu}^a,
&\hspace{1cm}c^a=\sqrt{Z_c} c_R^a\\
\bar c^a=\sqrt{Z_c} \bar c_R^a,
&\hspace{1cm}\xi_0=\frac{Z_A}{Z_\xi} \xi_R,\\
m^2_0=Z_m m^2_R,
&\hspace{1cm}g_0=Z_g g_R.
\end{array}
\end{equation}
Previous parameters (\ref{previous1},\ref{previous2}) can be expressed in terms of bare ones and renormalization factors (\ref{renfact}) by
\begin{equation}
\begin{array}{ll}
\xi=\frac{1}{Z Z_c}\xi_0=\frac{Z_A}{Z_\xi^2}\xi_R,
&\hspace{1cm}Z_\xi=Z Z_c\\
Z_m=\frac{Z Z_\xi}{Z_A},
&\hspace{1cm}g=\frac{Z_\xi}{Z_g Z_c \sqrt{Z_A}} g_0=\frac{Z_\xi}{Z_c \sqrt{Z_A}} g_R,
\end{array}
\end{equation}
where the relation to renormalized parameters has also been given.
One concludes that renormalization factors are not independent. They must satisfy the non-renormalization theorem
\begin{equation}
\label{tnr1}
Z_m=\frac{Z_\xi^2}{Z_c Z_A}.
\end{equation}
Renormalization factors for the Curci-Ferrari model have been calculated up to three loops in the $\overline{MS}$ scheme by Gracey \cite{Gracey02} and one can verify explicitly that these renormalization factors verify (\ref{tnr1}).
In fact, in the recent article \cite{Browne06} this relation is observed in the three loops perturbative series but an all order argument and the associated symmetry is absent.

Now, one can consider linear terms in $\bar L$ in the identity (\ref{STBRST}).
Terms proportional to $f^{abc}\bar L^a L^b c^c$ give another non-renormalization theorem:
\begin{equation}
\label{tnr2}
Z_g=\frac{Z_\xi^2}{\sqrt{Z_A} Z_c} .
\end{equation}
Notice that the Landau gauge is the intersection of usual covariant linear gauges and the Curci-Ferrari gauge.
Then, in the Landau gauge limit $\xi\to 0$, the covariant linear-gauge non-renormalization theorem of the gauge parameter is valid
$Z_\xi \sim 1$. Relations (\ref{tnr1}) and (\ref{tnr2}) then become usual Landau gauge's non-renormalization theorems \cite{Taylor71,Dudal02}.

Moreover, terms proportional to $\bar L^a \partial_\mu K_\mu^a$ implies that factor $\gamma$ is compatible with
multiplicative renormalizability:
\begin{equation}
\gamma= \frac{Z_\xi^2}{Z_c}.
\end{equation}
\section{Conclusions and perspectives}

In this paper some particular form of Slavnov-Taylor identities valid for the Curci-Ferrari model have been deduced.
They allow to deduce two non renormalization theorems that reduce the number of independent renormalization factors
in this model from five to three. These relations have been tested in the already known $\overline{MS}$ three loops calculation of these renormalization factors.
One of the two identities implies that the mass renormalization is fixed by the renormalization of dimensionless coupling in $d=4$.
This simplifies considerably the corresponding Callan-Symanzik equation, whose analysis is actually in progress
\cite{Binosi}.

Even if in this paper this particular Slavnov-Taylor identity has only been exploited in order to study renormalization properties of the Curci-Ferrari model, it can also be used to simplify the analysis of finite parts of vertices of the model, mainly in order to develop non-perturbative renormalizations schemes in Schwinger-Dyson or similar contexts.

\section{Acknowledgments}
I would like to thank D. Binosi,
R. M\'endez-Galain, M. Tissier and particularly J.P. Blaizot for useful discussions and reading
a draft of the present paper. I would like to thank L. von Smekal for
pointing out a missprint in a previous version of this manuscript.

\bibliographystyle{unsrt}

\begin{thebibliography}{10}

\bibitem{Alkofer00}
R.~Alkofer and L.~von Smekal,
Phys.\ Rept.\  {\bf 353} (2001) 281
[arXiv:hep-ph/0007355].

\bibitem{Zwanziger01}
D.~Zwanziger,
Phys.\ Rev.\ D {\bf 65} (2002) 094039
[arXiv:hep-th/0109224].

\bibitem{Aguilar01}
A.~C.~Aguilar, A.~Mihara and A.~A.~Natale,
Phys.\ Rev.\ D {\bf 65} (2002) 054011
[arXiv:hep-ph/0109223].

\bibitem{Zwanziger02}
D.~Zwanziger,
Phys.\ Rev.\ D {\bf 67} (2003) 105001
[arXiv:hep-th/0206053].

\bibitem{Alkofer02b}
R.~Alkofer, C.~S.~Fischer and L.~von Smekal,
Acta Phys.\ Slov.\  {\bf 52} (2002) 191
[arXiv:hep-ph/0205125].

\bibitem{Fischer02}
C.~S.~Fischer and R.~Alkofer,
Phys.\ Lett.\ B {\bf 536} (2002) 177
[arXiv:hep-ph/0202202].

\bibitem{Fischer02b}
C.~S.~Fischer, R.~Alkofer and H.~Reinhardt,
Phys.\ Rev.\ D {\bf 65} (2002) 094008
[arXiv:hep-ph/0202195].

\bibitem{Aguilar02}
A.~C.~Aguilar, A.~A.~Natale and P.~S.~Rodrigues da Silva,
Phys.\ Rev.\ Lett.\  {\bf 90} (2003) 152001
[arXiv:hep-ph/0212105].

\bibitem{Zwanziger03}
D.~Zwanziger,
Phys.\ Rev.\ D {\bf 69} (2004) 016002
[arXiv:hep-ph/0303028].

\bibitem{Alkofer03}
R.~Alkofer, W.~Detmold, C.~S.~Fischer and P.~Maris,
Phys.\ Rev.\ D {\bf 70} (2004) 014014
[arXiv:hep-ph/0309077].

\bibitem{Bloch03}
J.~C.~R.~Bloch, A.~Cucchieri, K.~Langfeld and T.~Mendes,
Nucl.\ Phys.\ B {\bf 687} (2004) 76
[arXiv:hep-lat/0312036].

\bibitem{Schleifenbaum04}
W.~Schleifenbaum, A.~Maas, J.~Wambach and R.~Alkofer,
arXiv:hep-ph/0411052.

\bibitem{Alkofer04}
R.~Alkofer, C.~S.~Fischer and F.~J.~Llanes-Estrada,
Phys.\ Lett.\ B {\bf 611} (2005) 279
[arXiv:hep-th/0412330].

\bibitem{Aguilar04}
A.~C.~Aguilar and A.~A.~Natale,
JHEP {\bf 0408} (2004) 057
[arXiv:hep-ph/0408254].

\bibitem{Pawlowski03}
J.~M.~Pawlowski, D.~F.~Litim, S.~Nedelko and L.~von Smekal,
Phys.\ Rev.\ Lett.\  {\bf 93} (2004) 152002
[arXiv:hep-th/0312324].

\bibitem{Fischer04}
C.~S.~Fischer and H.~Gies,
JHEP {\bf 0410} (2004) 048
[arXiv:hep-ph/0408089].

\bibitem{Becirevic99b}
D.~Becirevic, P.~Boucaud, J.~P.~Leroy, J.~Micheli, O.~Pene, J.~Rodriguez-Quintero and C.~Roiesnel,
Phys.\ Rev.\ D {\bf 60} (1999) 094509
[arXiv:hep-ph/9903364].

\bibitem{Becirevic99}
D.~Becirevic, P.~Boucaud, J.~P.~Leroy, J.~Micheli, O.~Pene, J.~Rodriguez-Quintero and C.~Roiesnel,
Phys.\ Rev.\ D {\bf 61} (2000) 114508
[arXiv:hep-ph/9910204].

\bibitem{Bakeev03}
T.~D.~Bakeev, E.~M.~Ilgenfritz, V.~K.~Mitrjushkin and M.~Mueller-Preussker,
Phys.\ Rev.\ D {\bf 69} (2004) 074507
[arXiv:hep-lat/0311041].

\bibitem{Cucchieri03}
A.~Cucchieri, T.~Mendes and A.~R.~Taurines,
Phys.\ Rev.\ D {\bf 67} (2003) 091502
[arXiv:hep-lat/0302022].

\bibitem{Cucchieri04}
A.~Cucchieri, T.~Mendes and A.~Mihara,
JHEP {\bf 0412} (2004) 012
[arXiv:hep-lat/0408034].

\bibitem{Sternbeck05}
A.~Sternbeck, E.~M.~Ilgenfritz, M.~Mueller-Preussker and A.~Schiller,
arXiv:hep-lat/0506007.

\bibitem{Boucaud05}
  P.~Boucaud {\it et al.},
  arXiv:hep-ph/0507104.

\bibitem{Bornyakov03}
V.~G.~Bornyakov, M.~N.~Chernodub, F.~V.~Gubarev, S.~M.~Morozov and M.~I.~Polikarpov,
Phys.\ Lett.\ B {\bf 559} (2003) 214
[arXiv:hep-lat/0302002].

\bibitem{Alkofer03b}
R.~Alkofer, C.~S.~Fischer, H.~Reinhardt and L.~von Smekal,
Phys.\ Rev.\ D {\bf 68} (2003) 045003
[arXiv:hep-th/0304134].

\bibitem{Szczepaniak03}
A.~P.~Szczepaniak,
Phys.\ Rev.\ D {\bf 69} (2004) 074031
[arXiv:hep-ph/0306030].

\bibitem{Zwanziger03b}
D.~Zwanziger,
Phys.\ Rev.\ D {\bf 70} (2004) 094034
[arXiv:hep-ph/0312254].

\bibitem{Taylor71}
J.~C.~Taylor,
Nucl.\ Phys.\ B {\bf 33} (1971) 436.

\bibitem{Wilson73}
  K.~G.~Wilson and J.~B.~Kogut,
  Phys.\ Rept.\  {\bf 12}, 75 (1974).

\bibitem{Polchinski83}
  J.~Polchinski,
  Nucl.\ Phys.\ B {\bf 231}, 269 (1984).

\bibitem{Wetterich92}
  C.~Wetterich,
  Phys.\ Lett.\ B {\bf 301}, 90 (1993).

\bibitem{Ellwanger93}
  U.~Ellwanger,
  Z.\ Phys.\ C {\bf 58}, 619 (1993).

\bibitem{Tetradis94}
  N.~Tetradis and C.~Wetterich,
  Nucl.\ Phys.\ B {\bf 422} (1994) 541.

\bibitem{Morris94b}
T.~R. Morris,
Int. J. Mod. Phys. A9 (1994) 2411--2450.

\bibitem{Morris94}
  T.~R.~Morris,
  Phys.\ Lett.\ B {\bf 329}, 241 (1994)
  [arXiv:hep-ph/9403340].

\bibitem{Berges00}
J. Berges, N. Tetradis and C. Wetterich,
   Phys. Rept. 363 (2002) 223--386.

\bibitem{Becchi03}
C.~Becchi, in Elementary Particles, Field Theory and Statistical Mechanics, eds.
M.~Bonini, G.~Marchesini and E.~Onofri, Parma University 1993.

\bibitem{Ellwanger94}
U.~Ellwanger,
Phys.\ Lett.\ B {\bf 335} (1994) 364
[arXiv:hep-th/9402077].

\bibitem{Bonini93}
M.~Bonini, M.~D'Attanasio and G.~Marchesini,
Nucl.\ Phys.\ B {\bf 421} (1994) 429
[arXiv:hep-th/9312114].

\bibitem{Bonini93b}
M.~Bonini, M.~D'Attanasio and G.~Marchesini,
Nucl.\ Phys.\ B {\bf 418} (1994) 81
[arXiv:hep-th/9307174].

\bibitem{Bonini94b}
M.~Bonini, M.~D'Attanasio and G.~Marchesini,
Phys.\ Lett.\ B {\bf 346} (1995) 87
[arXiv:hep-th/9412195].

\bibitem{Bonini94}
M.~Bonini, M.~D'Attanasio and G.~Marchesini,
Nucl.\ Phys.\ B {\bf 437} (1995) 163
[arXiv:hep-th/9410138].

\bibitem{Ellwanger95}
U.~Ellwanger, M.~Hirsch and A.~Weber,
Z.\ Phys.\ C {\bf 69} (1996) 687
[arXiv:hep-th/9506019].

\bibitem{Ellwanger96}
U.~Ellwanger, M.~Hirsch and A.~Weber,
Eur.\ Phys.\ J.\ C {\bf 1} (1998) 563
[arXiv:hep-ph/9606468].

\bibitem{Slavnov72}
A.~A.~Slavnov,
Theor.\ Math.\ Phys.\  {\bf 10}, 99 (1972)
[Teor.\ Mat.\ Fiz.\  {\bf 10}, 153 (1972)].

\bibitem{Morris99}
  T.~R.~Morris,
  Nucl.\ Phys.\ B {\bf 573}, 97 (2000)
  [arXiv:hep-th/9910058].

\bibitem{Morris00}
  T.~R.~Morris,
  JHEP {\bf 0012}, 012 (2000)
  [arXiv:hep-th/0006064].

\bibitem{Arnone01}
  S.~Arnone, Y.~A.~Kubyshin, T.~R.~Morris and J.~F.~Tighe,
  Int.\ J.\ Mod.\ Phys.\ A {\bf 17}, 2283 (2002)
  [arXiv:hep-th/0106258].

\bibitem{Arnone02}
  S.~Arnone, A.~Gatti and T.~R.~Morris,
  Phys.\ Rev.\ D {\bf 67}, 085003 (2003)
  [arXiv:hep-th/0209162].

\bibitem{Morris05}
  T.~R.~Morris and O.~J.~Rosten,
  Phys.\ Rev.\  D {\bf 73}, 065003 (2006)
  [arXiv:hep-th/0508026].

\bibitem{Rosten05}
  O.~J.~Rosten,
  J.\ Phys.\ A  {\bf 39}, 8699 (2006)
  [arXiv:hep-th/0507166].

\bibitem{Morris06}
  T.~R.~Morris and O.~J.~Rosten,
  J.\ Phys.\ A {\bf 39}, 11657 (2006)
  [arXiv:hep-th/0606189].

\bibitem{Callan70}
  C.~G.~.~Callan,
  Phys.\ Rev.\ D {\bf 2}, 1541 (1970).

\bibitem{Symanzik70}
  K.~Symanzik,
  Commun.\ Math.\ Phys.\  {\bf 18}, 227 (1970).

\bibitem{Curci76}
  G.~Curci and R.~Ferrari,
  Nuovo Cim.\ A {\bf 32}, 151 (1976).

\bibitem{Curci76b}
  G.~Curci and R.~Ferrari,
  Nuovo Cim.\ A {\bf 35}, 1 (1976)
  [Erratum-ibid.\ A {\bf 47}, 555 (1978)].

\bibitem{deBoer95}
  J.~de Boer, K.~Skenderis, P.~van Nieuwenhuizen and A.~Waldron,
  Phys.\ Lett.\ B {\bf 367}, 175 (1996)
  [arXiv:hep-th/9510167].

\bibitem{Ojima81}
  I.~Ojima,
  Z.\ Phys.\  C {\bf 13} (1982) 173.

\bibitem{Becchi74}
  C.~Becchi, A.~Rouet and R.~Stora,
  Commun.\ Math.\ Phys.\  {\bf 42}, 127 (1975).
  
\bibitem{Becchi75}
C.~Becchi, A.~Rouet and R.~Stora,
Annals Phys.\  {\bf 98} (1976) 287.

\bibitem{Tyutin75}
I.~V.~Tyutin, LEBEDEV-75-39.

\bibitem{Neuberger86}
  H.~Neuberger,
  Phys.\ Lett.\ B {\bf 175}, 69 (1986).

\bibitem{Neuberger86b}
  H.~Neuberger,
  Phys.\ Lett.\ B {\bf 183}, 337 (1987).

\bibitem{Gribov77}
V.~N.~Gribov,
Nucl.\ Phys.\ B {\bf 139} (1978) 1.

\bibitem{Kalloniatis05}
A.~C.~Kalloniatis, L.~von Smekal and A.~G.~Williams,
Phys.\ Lett.\ B {\bf 609} (2005) 424
[arXiv:hep-lat/0501016].

\bibitem{Ghiotti06}
  M.~Ghiotti, L.~von Smekal and A.~G.~Williams,
  AIP Conf.\ Proc.\  {\bf 892}, 180 (2007)
  [arXiv:hep-th/0611058].

\bibitem{Periwal95}
  V.~Periwal,
  arXiv:hep-th/9509084.

\bibitem{Dudal02}
  D.~Dudal, H.~Verschelde and S.~P.~Sorella,
  Phys.\ Lett.\ B {\bf 555}, 126 (2003)
  [arXiv:hep-th/0212182].

\bibitem{Delduc89}
  F.~Delduc and S.~P.~Sorella,
  Phys.\ Lett.\ B {\bf 231}, 408 (1989).

\bibitem{Alvarez-Gaume81}
  L.~Alvarez-Gaume and L.~Baulieu,
  Nucl.\ Phys.\  B {\bf 212}, 255 (1983).

\bibitem{Weinberg96}
S.~Weinberg,
``The quantum theory of fields. Vol. 2: Modern applications, Cambridge, UK: Univ. Pr. (1996) 489 p''

\bibitem{Nakanishi90}
N.~Nakanishi and I.~Ojima, World Sci. \ Lect. \ Notes Phys. \ {\bf 27} (1990) 1.

\bibitem{Gracey02}
  J.~A.~Gracey,
  Phys.\ Lett.\ B {\bf 552}, 101 (2003)
  [arXiv:hep-th/0211144].

\bibitem{Browne06}
  R.~E.~Browne {\it et al.},
  J.\ Phys.\ A {\bf 39}, 7889 (2006)
  [arXiv:hep-th/0602291].

\bibitem{Binosi}
D.~Binosi, J.P.~Blaizot, R.~M\'endez-Galain and N.~Wschebor, in preparation.

\end{thebibliography}

\end{document}